# Ohmic contacts on n-type and p-type cubic silicon carbide (3C-SiC) grown on Silicon

M. Spera[1,2,3], G. Greco[1], R. Lo Nigro[1], C. Bongiorno[1], F. Giannazzo[1], M. Zielinski[4], F. La Via[1], F. Roccaforte[1,*]

[1] *Consiglio Nazionale delle Ricerche – Istituto per la Microelettronica e Microsistemi (CNR-IMM), Strada VIII, n.5 Zona Industriale, I-95121 Catania, Italy*
[2] *Department of Physics and Astronomy, University of Catania, Via Santa Sofia, 64, 95123, Catania, Italy*
[3] *Department of Physics and Chemistry, University of Palermo, Via Archirafi, 36, 90123, Palermo, Italy*
4 *NOVASiC, Savoie Technolac, BP267, F-73375 Le Bourget-du-Lac Cedex, France*

*(\*) Corresponding author:  fabrizio.roccaforte@imm.cnr.it*

**Abstract**

This paper is a report on Ohmic contacts on n-type and p-type type cubic silicon carbide (3C-SiC) layers grown on silicon substrates. In particular, the morphological, electrical and structural properties of annealed Ni and Ti/Al/Ni contacts has been studied employing several characterization techniques. Ni films annealed at 950°C form Ohmic contacts on moderately n-type doped 3C-SiC ($N_D \sim 1\times10^{17}cm^{-3}$), with a specific contact resistance of $3.7\times10^{-3}$ $\Omega cm^2$. The main phase formed upon annealing in this contact was nickel silicide ($Ni_2Si$), with randomly dispersed carbon in the reacted layer. In the case of a p-type 3C-SiC with a high doping level ($N_A \sim 5\times10^{19}cm^{-3}$), Ti/Al/Ni contacts were preferable to Ni ones, as they gave much lower values of the specific contact resistance ($1.8 \times10^{-5}$ $\Omega cm^2$). Here, an $Al_3Ni_2$ layer was formed in the uppermost part of the contact, while TiC was detected at the interface. For this system, a temperature dependent electrical characterization allowed to establish that the thermionic field emission rules the current transport at the interface. All these results can be useful for the further development of a devices technology based on the 3C-SiC polytype.

## 1. Introduction

Today, silicon carbide (SiC) devices research and development is mainly based on the hexagonal polytype (4H-SiC), due to the availability of large area wafers (up to 150mm) of good crystalline and electronic quality [1,2]. However, it has been often reported that the cubic polytype (3C-SiC) can give some advantages with respect to 4H-SiC. In fact, due to the narrower band gap (~2.2eV) of this polytype [3], the $SiO_2$/3C-SiC interface is expected to have a lower density of interface traps with respect to the $SiO_2$/4H-SiC system [4,5]. Consequently, high inversion channel mobility values can be achieved in metal-oxide-semiconductor field effect transistors (MOSFETs) [6,7]. Moreover, 3C-SiC is the only polytype that can be heteroepitaxially grown on large diameter silicon (Si) substrates. This aspect makes this semiconductor a very promising material for the fabrication of power devices on a large scale and at a low cost.

In spite of these potential advatages, the application of 3C-SiC in power devices technology remains still limited by the material quality, which in turn is strongly related to the growth techniques [8]. In this scenario, also the electrical behavior of the fundamental devices building blocks – such as metal/semiconductor contacts – can be significantly influenced by the quality of the 3C-SiC material (e.g., defect density, surface roughness, surface preparation, etc.) [9,10,11].

Ohmic contacts are fundamental bricks both for diodes and MOSFETs. In the past, Ohmic contacts formed using annealed Nickel- and Titanium-based metallic layers have been investigated on n-type heavily doped 3C-SiC layers, grown on different substrates [9,12,13]. Very recently, Ohmic contacts were obtained on heavily doped (degenerate) n-type implanted 3C-SiC without annealing and applied to 3C-SiC MOSFETs fabrication [14]. However, only a limited work has been reported on moderately doped n-type 3C-SiC and on p-type doped 3C-SiC [15,16]. In this context, a renewed interest for 3C-SiC is presently associated to the running European research programs [17], which stimulate further investigations on the device processing on this polytype.

This work reports on the fabrication and characterization of Ohmic contacts on moderately doped n-type and heavily doped p-type 3C-SiC layers grown on Si. The use of several morphological, structural and electrical analyses allowed to get information on the key factors for Ohmic contact formation based on annealed Ni and Ti/Al/Ni layers on 3C-SiC materials.

## 2. Experimental

The material used in this work consisted in 3C-SiC layers, grown in hot-wall chemical vapor deposition (CVD) reactors on 4-inches Si(100) substrates. Typically, Silane ($SiH_4$) and ethylene ($C_2H_4$) or propane ($C_3H_8$) were used as supplier gases for Si and C, respectively [18,19]. For this investigation on Ohmic contacts, two different samples were grown: moderately doped n-type 3C.SiC layers ($N_D \sim 10^{17}$ cm$^{-3}$) and heavily doped p-type 3C-SiC layers ($N_A \sim 5\times10^{19}$ cm$^{-3}$). The investigated metal systems were $Ni_{(100nm)}$ films and $Ti_{(70nm)}/Al_{(200nm)}/Ni_{(50nm)}$ multilayers, which have been deposited by sputtering and annealed at 950°C in Ar atmosphere for 60s in a Jipelec Jet First furnace. Circular TLM (C-TLM) structures [20] have been defined by optical lithography and metal lift-off, and were used for the estimation of the specific contact resistance of the contacts. Atomic force microscopy (AFM) was employed to monitor the surface morphology of the contacts, using a PSIA XE-150 microscope. X-ray Diffraction (XRD) analysis in grazing mode was performed on unpatterned samples using a Bruker-AXS D5005 θ-θ diffractometer, to identify the main phases formed during annealing. Finally, Transmission Electron Microscopy (TEM) analyses have been carried out on cross-sectional samples using a 200 kV JEOL 2010 F microscope, equipped with Energy Filtered Transmission Electron Microscopy (EFTEM).

## 3. Results and Discussion

First, Ohmic contacts were studied on the moderately doped n-type 3C-SiC material.
Figs. 1a and 1b show the AFM topographies collected on the bare 3C-SiC surface and on the Ni-contact surface annealed at 950°C. As can be seen, the as-grown 3C-SiC material is characterized by irregular surface features, determining a high surface roughness with root mean square (RMS) values of 18.6 nm. This roughness is likely associated to planar defects, such as stacking faults and twins, as well as anti-phase boundaries characterizing the heteroepitaxial 3C-SiC layers [18]. On the other hand, a higher RMS is measured on the annealed contact (33.1 nm), likely associated to a solid state reaction occurring upon annealing.

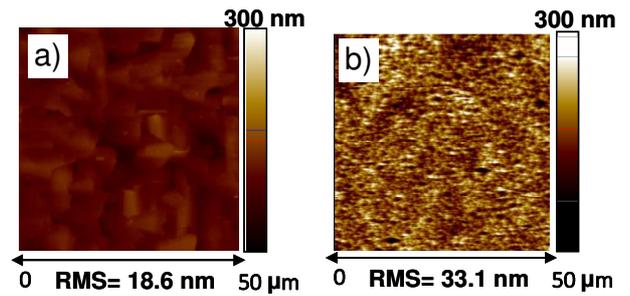

**Fig. 1.** AFM images of the bare n-type 3C-SiC sample surface (a) and of the Ni contact after annealing at 950°C (b).

Fig. 2 reports the I-V characteristics acquired on adjacent C-TLM pads for the Ni contacts annealed at 950°C. From a linear fit of the plot of the total resistance $R_{TOT}$ as a function of the C-TLM pad distance (see inset of Fig. 2) it was possible to extract a specific contact resistance of $3.7\times10^{-3}$ Ωcm$^2$.

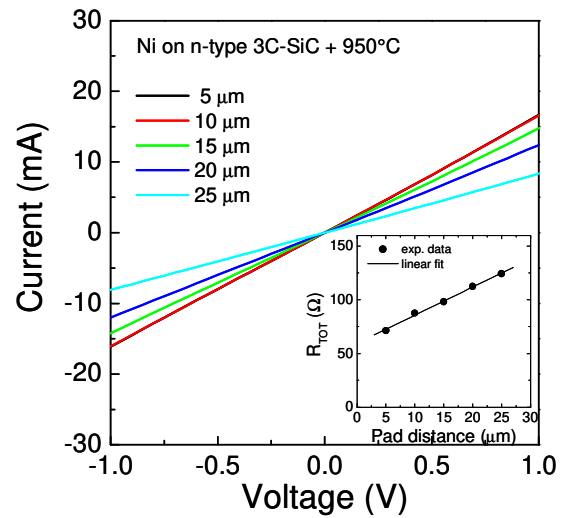

**Fig. 2.** I-V curves acquired on C-TLM structures at different distances for annealed (950°C) Ni contacts on moderately doped n-type 3C-SiC. The inset reports a plot of the total resistance $R_{TOT}$ as a function of the pad distance, used for the extraction of the specific contact resistance.

This value is reasonably good, considering that the material has only a moderate doping level.
Fig. 3a shows the XRD patterns acquired on the Ni/3C-SiC sample after annealing at 950 °C. The XRD peaks indicate the formation of the nickel silicide phase ($Ni_2Si$). Moreover, cross section TEM analysis of the sample shows an irregular interface (originating from the pristine surface morphology of the material) and the disordered distribution of carbon clusters inside the reacted metal layer. Hence, the formation of the silicide ($Ni_2Si$) is a key factor for the achievement of a Ohmic behavior.

Afterwards, Ohmic contacts were formed on the heavily doped p-type 3C-SiC layer. In particular, first the same Ni contact already used for the n-type material has been tested also for the heavily doped p-type samples. In this case, in spite of the high acceptor concentration level, the specific contact resistance obtained after Ni$_2$Si formation was 4.9 ×10$^{-3}$ Ωcm$^2$, i.e., in the same order of magnitude of that obtained in the moderately doped n-type material. This result confirms that the formation of a good Ohmic contact on a p-type wide band gap semiconductor represents a challenging issue [21]. In order to improve the Ohmic behavior of the contacts on the p-type material, a low work function metal can be used [21]. For this purpose, a Ti/Al/Ni stack has been investigated, being the low work function metal (Ti) in direct contact with the 3C-SiC.

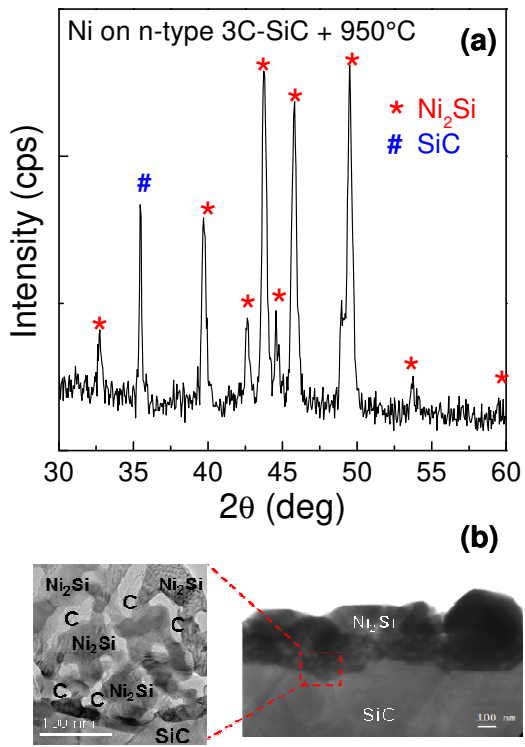
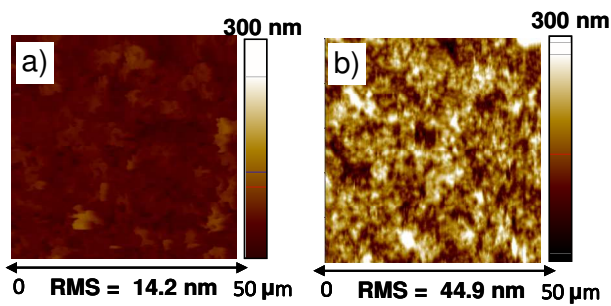

**Fig. 3.** (a) XRD diffraction patterns and (b) cross section TEM micrographs of Ni/3C-SiC samples after annealing at 950°C.

Fig. 4 shows the AFM images of the surface of the p-type 3C-SiC material and that of the Ti/Al/Ni contact formed on it after annealing at 950°C. The RMS values of these surfaces were 14.2 nm and 44.9 nm, respectively.

**Fig. 4.** AFM images of the bare p-type 3C-SiC sample surface (a) and of the Ti/Al/Ni contact after annealing at 950°C.

Clearly, the annealed Ti/Al/Ni layers exhibit a higher RMS with respect to the Ni$_2$Si ones. In this case, however, the electrical analyses of the C-TLM structures (see Fig. 5 and inset) demonstrated a significantly lower value of the specific contact resistance, i.e., 1.8 ×10$^{-5}$ Ωcm$^2$.

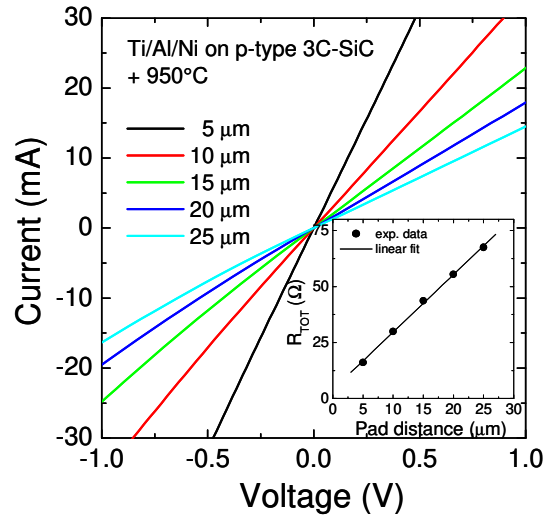

**Fig. 5.** I-V curves acquired on C-TLM structures at different distances for annealed (950°C) Ti/Al/Ni contacts on heavily doped p-type 3C-SiC. The inset reports a plot of the total resistance $R_{TOT}$ as a function of the pad distance, used for the extraction of the specific contact resistance.

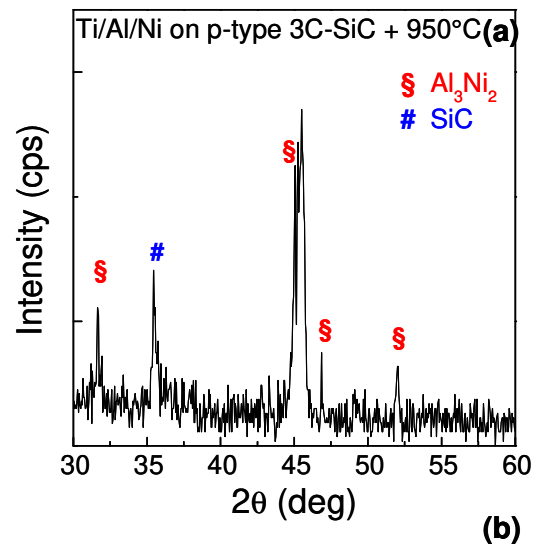
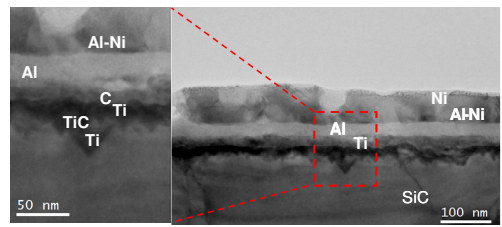

**Fig. 6.** (a) XRD diffraction patterns and (b) cross section TEM micrographs of Ti/Al/Ni/3C-SiC samples after annealing at 950°C.

The structural analysis of the samples performed by XRD (Fig. 6a) revealed the presence of the $Al_3Ni_2$ phase, which is typically localized in the uppermost part of the reacted stack (see TEM image in Fig. 6b), similarly to the behavior of the same multilayer on 4H-SiC [22]. Moreover, cross section TEM (Fig. 6b) also demonstrated the presence of a thin TiC layer close to the interface with 3C-SiC, which is likely responsible for the Ohmic behavior of the contact.

The ternary phase $Ti_3SiC_2$, sometimes detected upon annealing of Ti/Al bilayers on 4H-SiC [23], was not observed in our Ti/Al/Ni system, i.e., in the presence of a Ni cap layer.

Finally, the temperature dependence of the specific contact resistance in Ti/Al/Ni contacts on the heavily doped p-type 3C-SiC was studied to get insights into the transport mechanism at the metal/semiconductor interface. In particular, as can be seen in Fig. 7, the values of $\rho_C$ slightly decreased with increasing the measurement temperature.

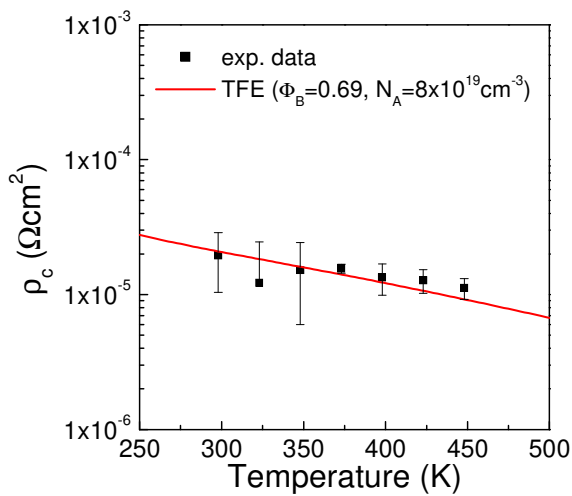

**Fig. 7.** Temperature dependence of the specific contact resistance of Ti/Al/Ni Ohmic contacts on p-type 3C-SiC. The continuous line is the fit of the experimental data with the TFE model.

The experimental $\rho_c$ data could be fitted assuming a thermionic field emission (TFE) mechanism [20]. From the fit with the TFE model of the experimental data (depicted as continuous line in Fig. 7) a barrier height of $\Phi_B=0.69$ eV and a carrier concentration $N_A=8\times10^{19}$ cm$^{-3}$ have been determined. The value of the barrier height is higher that the one measured on 4H-SiC (0.56 eV) [22]. Hence, although a lower barrier could be expected for the cubic polytype, this results suggests that the interface composition and roughness of the reacted layer on 3C-SiC plays a role in the formation of the Ohmic contact and of its barrier.

### 4. Conclusion

In summary, this paper reported on the fabrication and characterization of Ohmic contacts on moderately doped n-type 3C-SiC and heavily doped p-type 3C-SiC, considering Ni and Ti/Al/Ni contacts. A Ohmic behavior could be observed after annealing at 950°C, with values of the specific contact resistance $\rho_c$ varying in the range ~ $10^{-3}$-$10^{-5}$ $\Omega cm^2$, depending on the 3C-SiC layer and on the metal. In general, annealed Ni contacts exhibited a better surface morphology than Ti/A/Ni ones. However, these latter are more suitable to form Ohmic contacts on p-type doped material, owing to the lower work function. The results can be useful for the future development of the devices technology based on 3C-SiC in different applications fields.

### Acknowledgements

This work has been supported by the European project Challenge (Call: H2020-NMBP-2016-2017, grant. agreement 720827).

The authors would like to thank P. Fiorenza for fruitful discussions and S. Di Franco for technical support during sample processing.